\documentclass{aa}
\usepackage{graphicx}
\begin{document}
   \title{The Optical afterglow of the not so dark GRB 021211}
   
   \author{S.B. Pandey\inst{1}, G.C. Anupama\inst{2}, R. Sagar\inst{1,2},
   D. Bhattacharya\inst{3},   A.J. Castro-Tirado\inst{4}, D.K. Sahu\inst{2,5}, Padmakar Parihar\inst{2,5} 
          and T.P. Prabhu\inst{2}}

   \institute{ 
   State Observatory, Manora peak, Naini Tal - 263129, Uttaranchal, India
   \and
   Indian Institute of Astrophysics, Bangalore -- 560 034, India
   \and
   Raman Research Institute, Bangalore -- 560 080, India
   \and
   Instituto de Astrof\'isica de Andaluc\'ia, P.O. Box 03004, E-18080, Granada, Spain
   \and
   Center for Research \& Education in Science \& Technology, Hosakote, Bangalore -- 562 114, India
}

   \authorrunning{S. B. Pandey et al.}
   \titlerunning{The optical afterglow of the not so dark GRB 021211}
   \offprints{S. B. Pandey, \\
           \email{shashi@upso.ernet.in}
            }   
   \date{Received ------ /accepted ---------}

  \abstract{
We determine Johnson $B,V$ and Cousins $R,I$ photometric CCD magnitudes for
the afterglow of GRB~021211 during the first night after the GRB trigger.  
The afterglow was very faint and would have been probably missed if no prompt
observations had been conducted.  A fraction of the so-called ``dark'' GRBs
may thus be just ``optically dim'' and require very deep imaging to be
detected. The early-time optical light curve reported by other observers
shows a prompt emission with properties similar to that of GRB~990123.  Following 
this, the afterglow emission from $\sim 11$~min to $\sim 35$~days after the burst 
is characterized by an overall power-law decay with a slope $1.1\pm0.01$ in the $R$ 
passband. We derive the value of spectral index in the optical to near-IR
region to be 0.6$\pm$0.2 during 0.13 to 0.8 day after the burst.
The flux decay constant and the spectral slope indicate that during the
first day after the burst, the optical band lies between the cooling 
frequency and the synchrotron maximum frequency of the afterglow. 

}

\maketitle
\section{Introduction}

A long duration burst, GRB 021211 ($\equiv$ H2493), triggered at 
11$^h$18$^m$34.$^s$03 UT on 11 December 2002 was detected by the High Energy 
Transient Explorer (HETE) FREGATE, WXM, and soft $X-$ray camera (SXC) 
instruments (Crew et al. 2003). It was also observed by ULYSSES and KONUS 
(Hurley et al. 2002). The burst had a duration of $\sim$ 2.3 seconds at higher 
energies (85 -- 400 keV) but a longer duration of about 8.5 seconds at lower energies 
(5 -- 10 keV) band. It had a fluence of about 1 and 2 $\mu$erg/cm$^2$ in the energy 
bands of 7 -- 30 keV and 30 -- 400 keV respectively. This indicates that GRB 021211 
is an ``$X -$ ray rich'' burst (Crew et al. 2003). The SXC coordinates of the burst 
reported by Crew et al. (2003) are $\alpha = 08^{\rm h} 09^{\rm m} 00^{\rm s}, 
\delta = +06^{\circ} 44^\prime 20^{\prime\prime}$ (J2000). Within the error 
circle of SXC, an optical afterglow (OA) of the GRB 021211 was discovered by 
Fox \& Price  (2002) at $\alpha = 08^{\rm h} 08^{\rm m} 59.^{\rm{s}}883,  
\delta = +06^{\circ} 43^\prime 37.^{\prime\prime}88$ (J2000).
The source was subsequently also identified in a number of images taken at 
$\sim$ 90, 108 and 143 seconds after the burst by robotic optical telescopes.  
Thus, GRB 021211 joins the group of GRB 990123 (Akerlof et al. 1999) and 
GRB 021004 (Fox et al. 2003b, Pandey et al. 2003) whose early optical emissions 
could be observed within few minutes of the trigger of the event. 
Spectroscopic observations by Della Valle et al. (2003)
indicate a redshift value of $z$ = 1.004$\pm$0.002 for the probable host galaxy
of GRB 021211. Fox et al. (2003a) report optical and near-IR observations of the 
GRB afterglow and find that at optical wavelengths, the GRB 021211 afterglow is 
significantly fainter than most of the known afterglows at an epoch of 
$\sim$ 1 day. The observed fluence in the 30 -- 400 keV energy band by Crew et al. 
(2003) together with the measured redshift $z$ = 1.004$\pm$0.002 (Della Valle et al. 2003) 
indicates an isotropic equivalent energy release $E_{{\rm iso},\gamma}\sim 6.1 
\times 10^{51}$~erg for $H_0$ = 65 km/s/Mpc in a $\Omega_0$ = 0.3 and $\Lambda_0$ = 
0.7 cosmological model. With a cosmological $K$-correction as in Bloom, Frail \& Sari 
(2001) the estimated isotropic-equivalent energy becomes $E_{{\rm iso},\gamma}\sim 
1.02\times 10^{52}$~erg, an order of magnitude lower than the corresponding 
estimate for GRB~990123 (Bloom, Frail \& Kulkarni 2003).

In this paper we present optical observations obtained during the temporal gap 
of the light curves presented by Della Valle et al. (2003a), Fox et al. (2003a) 
and Li et al. (2003) using secure photometric calibrations.

\section {Observations and data reduction}
The broad band Johnson $BV$ and Cousins $RI$ photometric observations of the OA were 
carried out on 11 December 2002 using the 104-cm Sampurnanand telescope of the 
State Observatory, Nainital and 2-m Himalayan Chandra Telescope (HCT) of the Indian 
Astronomical Observatory (IAO), Hanle. At Nainital, one pixel of the
2048 $\times$ 2048 pixel$^{2}$ size CCD chip corresponds to a square of
0.$^{''}$38 side, and the entire chip covers a field of $\sim 13^{\prime}\times 
13^{\prime}$ on the sky. The gain and read out noise of the CCD camera are 10 
$e^-/ADU$ and 5.3 $e^-$ respectively. At Hanle, one pixel corresponds to a 
square of 0.$^{''}$3 side, and the entire chip covers a field of $\sim 10^{\prime} 
\times 10^{\prime}$ on the sky. It has a read out noise of 4.95 $e^-$ and gain is 
1.23 $e^-/ADU$. From Nainital, the CCD $BVRI$ observations of the OA field along 
with Landolt (1992) standard SA 98 region were obtained on 26/27 December 2002 
for photometric calibrations during good photometric sky conditions. During the 
observing run, several twilight flat field and bias frames were also obtained for 
the CCD calibrations.

ESO MIDAS, NOAO IRAF and DAOPHOT softwares were used to process the CCD frames 
in a standard way. The photometric calibrations derived using the six standards 
of the SA 98 region with color $0.61 < (V-I) < 2.14$ and brightness 
$13.1 < V < 16.3$ are: 

\noindent $b_{\rm{CCD}} = B - (0.036\pm0.01) (B-V) + (4.75\pm0.01) $  \\
        $v_{\rm{CCD}} = V - (0.027\pm0.01) (B-V) + (4.30\pm0.01) $  \\
        $r_{\rm{CCD}} = R -(0.004\pm0.01) (V-I) + (4.23\pm0.02) $  \\
        $i_{\rm{CCD}} = I - (0.064\pm0.01) (V-I)  + (4.73\pm0.01) $  \\

where $BVRI$ are standard magnitudes and $b_{\rm{CCD}}, v_{\rm{CCD}}, 
r_{\rm{CCD}} $ and
$i_{\rm{CCD}}$ represent the instrumental aperture magnitudes normalized for 
1 second of exposure time and corrected for atmospheric extinction 
coefficients determined from the Nainital observations of SA 98 bright stars. The 
values are 0.27, 0.17, 0.11 and 0.10 mag at the zenith in $B,V,R$ and $I$ filters 
respectively on the night of 26/27 December 2002. The errors in the colour 
coefficients and zero points are obtained by fitting least square linear 
regressions to the data points. Using the above calibrations, $BVRI$ photometric
magnitudes of 10 secondary standard stars are determined in the GRB 021211
field and their average values are listed in Table 1. The $(X,Y)$ CCD
pixels are used to convert coordinates into equatorial coordinates $\alpha_{2000}, 
\delta_{2000}$ values using the astrometric positions given by Henden (2002). 
All the secondary stars have been observed seven times in a filter and have internal 
photometric accuracy better than 0.01 mag. 
A comparison of present magnitudes of the secondary stars
with those given by Henden (2002) values yields zero-point differences of 
$0.04\pm0.01, 0.01\pm0.02, 0.01\pm0.02$ and $0.00\pm0.02$ mag in $V, (B-V),
(V-R)$ and $(V-I)$ respectively. Zero point difference is thus significant
in $V$, however these numbers can be accounted
in terms of the errors present in the zero point determination of the two
photometries. There is no colour dependence in the photometric differences.
These demonstrate that the photometric calibrations used in the present work are secure.

\begin{table}
{\bf Table 1}
{The identification number(ID), $(\alpha , \delta)$ for epoch 2000,
standard $V, (B-V), (V-R)$ and $(R-I)$ photometric magnitudes of the secondary 
standards in the GRB 021211 region.}


\scriptsize
\begin{tabular}{ccc cc ccl} \hline
ID & $\alpha_{2000}$ & $\delta_{2000}$ & $V$& $(B-V)$ & $(V-R)$ & $(V-I)$
 \\
    & (h m s) & (deg m s) & (mag) & (mag) & (mag) & (mag)  \\ \hline
  1&08 08 56&06 42 53& 16.81&  1.21&  0.73&  1.29 \\
  2&08 08 57&06 43 35& 17.01&  0.65&  0.40&  0.71 \\
  3&08 09 00&06 43 52& 15.43&  0.43&  0.30&  0.55 \\
  4&08 09 00&06 43 03& 14.25&  0.90&  0.53&  0.96 \\
  5&08 09 01&06 48 26& 14.99&  0.39&  0.26&  0.48 \\
  6&08 09 00&06 47 47& 15.61&  0.87&  0.50&  0.86 \\
  7&08 09 03&06 47 26& 13.90&  1.28&  0.69&  1.27 \\
  8&08 09 04&06 46 40& 16.60&  0.53&  0.35&  0.65 \\
  9&08 08 57&06 46 03& 14.70&  0.72&  0.44&  0.75 \\
 10&08 09 05&06 45 45& 15.34&  1.05&  0.67&  1.21 \\
\hline
\end{tabular}
\end{table}


Several short exposures up to a maximum of 30 min were generally given
while imaging the OA (see Table 2). In order to improve the signal-to-noise
ratio of the OA, the data have been binned in $2 \times 2$ pixel$^2$ and also
several bias corrected and flat-fielded CCD images of OA field
are co-added in the same filter, when found necessary. From these images,
profile-fitting magnitudes are determined using DAOPHOT software due to the
presence of bright star near the OT. The profile
magnitudes have been converted to aperture (about 5 arcsec) magnitudes using
aperture growth curve determined from well isolated secondary standards. They are
differentially calibrated using the secondary standards listed in Table 1 and
the values derived in this way are given in Table 2.


\begin{table}[ht]
{\bf Table 2}
{CCD BVRI broad band optical photometric observations of
the GRB 021004 afterglow. At Hanle, 2-m HCT was
used while at Nainital, 104-cm Sampurnanand optical telescope was used.}

\scriptsize



\begin{tabular}{cccll} \hline
Date (UT) & Magnitude & Exposure time& Passband & Telescope  \\
2002 December& (mag)&(Sec)&& \\   \hline

11.9347&22.7$\pm$0.14&2$\times$1200&B&104-cm\\ \\

11.8264&22.3$\pm$0.20&1800&V&104-cm\\
11.9632&$>$22.6&1800&V&104-cm\\ \\

11.7549&21.9$\pm$0.21&2$\times$900&R&104-cm\\
11.7792&21.9$\pm$0.16&2$\times$900&R&104-cm\\
11.8640&22.1$\pm$0.18&600&R&HCT\\
11.8730&22.1$\pm$0.18&600&R&HCT\\
11.8822&22.1$\pm$0.14&600&R&HCT\\
11.8914&22.4$\pm$0.24&600&R&HCT\\
11.8993&22.5$\pm$0.21&2$\times$1800&R&104-cm\\
11.9318&22.2$\pm$0.18&2$\times$600&R&HCT\\ \\

11.8028&$>$21.4&2$\times$900&I&104-cm\\

\hline
\end{tabular}
\end{table}

\section {Results}

\subsection {$R$ band photometric light curve}

In Fig. 1, we plot the temporal evolution of our $R$ band GRB 021211 afterglow 
measurements along with those published by Della Valle et al. (2003a), Fox et al. 
(2003a), Fruchter et al. (2002), Levan et al. (2002), Li et al. (2003), McLeod et 
al. (2002), Park et al. (2002) and Wozniak et al. (2002) after correcting for the 
host galaxy contribution as described in the next paragraph. 
We also make use of the published photometric measurements which could 
be converted on the present photometric scales using secondary stars listed in table 1. 
Fig. 1 also shows the $R$ band light curves of GRB 990123 and GRB 021004.
Early time observations of GRB 990123 ($\Delta t <$ 7~min) and GRB
021211 ($\Delta t <$ 11min) can be well explained in terms of reverse shock
emission, taking into account that GRB 021211 $\sim$ 4 mag fainter than GRB 990123
as noticed by Li et al. (2003) too. The GRB 021004 
early time ($\Delta t <$ 19~min) optical observations show unexpectedly shallower 
flux decay than that of GRB 990123 and GRB 021211 so reverse shock explanation 
can be ruled out either for homogeneous or for inhomogeneous environments 
(Chevalier \& Li 2000, Fox et al. 2003b).

The flux decay of the GRB 021211 OA, at times $>$ 11 min after the burst can be well 
characterized by a single power law decay plus a constant flux $F_{\rm{host}}$, component 
for the underlying host galaxy and can be written as

\begin{equation}
F(t) = const.\times t^{-\alpha} + F_{\rm{host}}
\end{equation}

\begin{figure}[h]
\centering
\includegraphics[height=9.0cm,width=9.0cm]{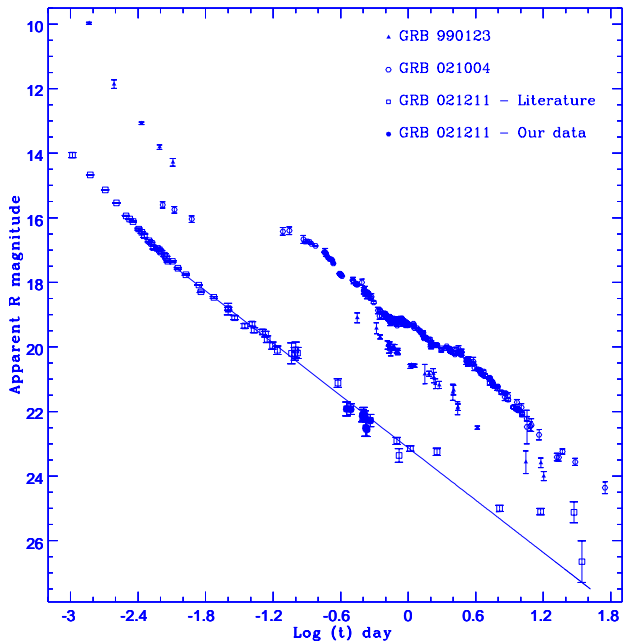}
\caption{\label{ospec} Light curve of the GRB 021211 OA in $R$ photometric passband.
Filled circles denote the present data, whereas empty squares are data taken from 
the references given in the main text. The solid line represents a single power-law 
fit for the flux decay after subtracting 25.16 mag, the fitted host galaxy contribution. 
GRB 990123 (Akerlof et al. 1999 \& Castro - Tirado et al. 1999)
and GRB 021004 (Pandey et al. 2003) $R$ band data are also plotted in the figure to
show the relative faintness of GRB 021211 as these GRBs have the early time 
optical observations. Time $t$ is measured from the GRB trigger. 
}
\end{figure}

\begin{figure}[h]
\centering
\includegraphics[height=9.0cm,width=9.0cm]{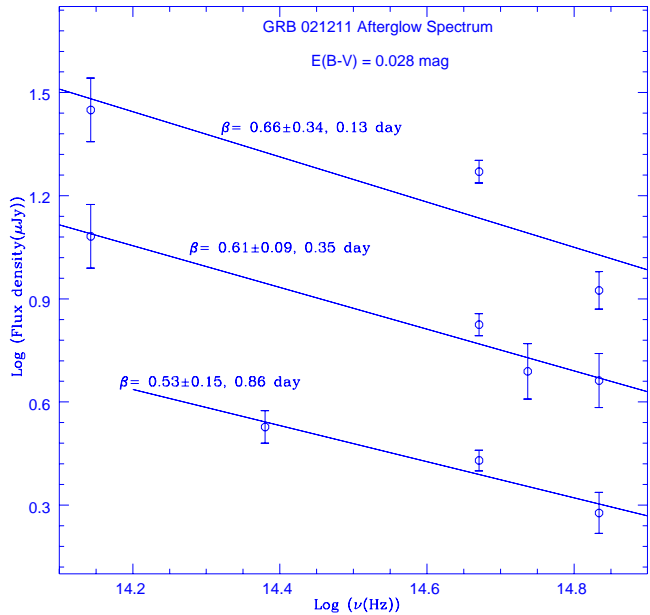}
\caption{\label{ospec} Optical-near IR spectrum of the GRB 021211 OA corrected 
for $E(B-V) = 0.028$ mag at $\sim$ 0.13, 0.35 and 0.86 days after the burst.}
\end{figure}

where $F(t)$ is total measured flux of the OT at time $t$ after the burst 
and $\alpha$ is the temporal flux decay index. We fitted the above function
using the least square regression method leaving $F_{\rm{host}}$ as a free parameter, 
including the late time VLT $R$ band observations (Della Valle et al. 2003a) but excluding 
the HST points (Fruchter et al. 2002). This yields $\alpha$ = 1.11$\pm$0.01, $\chi^2$ 
per DOF = 3.73 with host galaxy $R$ = 25.16$\pm$0.05 mag, which is
consistent with the value given by Della Valle et al. (2003a). 
The observations presented here follow the fitted light curve 
well and fill the existing temporal gap in the published data (see Fig 1).
The least square linear regression to the host galaxy contribution subtracted
data including HST points yields $\alpha$ = 1.08$\pm$0.01 with $\chi^2$ per 
DOF = 5.48. The larger values of $\chi^2$ are mainly due to two deviant
points (see Fig 1) towards later epoch. As argued by Della Valle et al. (2003a), 
these deviations most probably arise due to the additional contribution from a 
supernova component, noticeable in their spectroscopic data.

The value of flux decay $\alpha$ agrees fairly well with that determined by 
Fox et al. (2003a) and Li et al. (2003). The very fast flux decay, within a day 
after the burst, of the GRB 021211 OA can be compared with those of   
GRB 000630 (Fynbo et al. 2001), GRB 020124 (Berger et al. 2002) and GRB 020322 
(Burud et al. 2002) afterglowsi, which had similar temporal flux decay slopes 
and
were detected at $\sim$ 23 mag in the $R$ passband, one day after the burst.

\subsection {Spectral energy distribution}

Fig. 2 shows the GRB 021211 afterglow optical to near-IR spectrum at three epochs:
$\Delta t$ = 0.13, 0.35 and 0.86 day using the present $BVR$ optical data and
the published $BRJK_s$ observations by Fox et al. (2003a) and  Bersier et
al. (2003). The epochs are selected according to the widest possible wavelength coverage. 
Where necessary, measurements are interpolated at a given wavelength. We used the 
reddening map provided by Schlegel, Finkbeiner \& Davis (1998) for estimating 
Galactic interstellar extinction towards the burst and found a small value of 
$E(B - V) = 0.028$ mag. We used the standard Galactic extinction reddening curve 
given by Mathis (1990) in converting apparent magnitudes 
into fluxes and used the effective wavelengths and normalizations by Fukugita et al. 
(1995) and Bessell \& Brett (1988), for $BVR$ and Epchtein et al. (1994) for 
$J$ and $K_s$. We corrected the data for Galactic extinction only as 
the intrinsic extinction contribution from the host galaxy is unknown. 
We describe the spectrum 
by a single power law: $F_{\nu}\propto\nu^{-\beta}$, where $F_{\nu}$ is the flux at 
frequency $\nu$ and $\beta$ is the spectral index. The values of $\beta$ at 0.13, 
0.35 and 0.86 days after the burst are $0.66\pm$0.34, $0.61\pm$0.09 and 0.53$\pm$0.15 
respectively. The corresponding values of $(B - R)$ are 1.2$\pm$0.2, 0.8$\pm$ 0.2 and 
0.8$\pm$0.2 mag respectively. Whereas the $(B - K_s)$ values are 3.4$\pm$0.3 and 
3.2$\pm$0.3 at $\Delta t$ = 0.13 and 0.35 day respectively. These values indicate 
that the spectral slope of GRB 021211 OA has not changed within 
a day after the burst and has a mean value of 0.6$\pm$0.2. 

\section {Discussions and Conclusions}

$BVRI$ optical observations of the GRB 021211 OA around 0.28 day after the
burst are presented. The optical light curve of GRB 021211 OA (Fig.1) at times
$>$ 11 min after the burst can be well explained in terms of a single power 
law with the underlying host galaxy of $R = 25.16\pm0.05$ mag. GRB 011211 optical 
afterglow is intrinsically faint when compared with those of GRB 990123
and GRB 021004. It was detected only due to prompt, early follow up. Otherwise 
it would have been classified as a ``dark GRB'' as it was fainter than 
$R$ = 23 mag, 
$\sim$ 1 day after the burst and, in general, the usual follow-up observations 
do not go that deep. It thus appears that GRB 021211 is the first example of an 
``optically dim'' burst for which early time (less than a few minutes after the
burst) observations are available. It is thus likely that many optically ``dark GRBs'' 
could just be ``optically dim'' afterglows with the reason behind
their non-detection being not only due to the high redshift and extinction due 
to host galaxy but also due to the OA being much fainter than those observed to 
date (Crew et al. 2003). So, GRB 021211 is an example to indicate that a 
fraction of the otherwise so-called ``dark GRBs" are ``not so dark''. Deeper and 
faster follow-up observations are required to detect them.

Our fitted $R$ band values of temporal flux decay $\alpha$ and derived optical-near 
IR spectral slope $\beta$ can be well understood in terms of simple spherical 
adiabatic case for the homogeneous medium (Sari, Piran, \& Narayan 1998) 
in which for $\nu < \nu_{c}$, $\alpha = 3\beta/2 = 3(p - 1)/4$.
The observed spectral slope $\beta$ = 0.6$\pm$0.2 thus yields $\alpha$ = 0.9$\pm$0.3 
, consistent with the observed value of $\alpha$ = 1.1$\pm$0.01 and the electron energy 
distribution index $p = 2.2\pm0.4$. These values indicate that the cooling frequency 
$\nu_c$ lies above the optical band.

{\it Acknowledgments}
This research has made use of data
obtained through the High Energy Astrophysics Science Archive Research Center
Online Service, provided by the NASA/Goddard Space Flight Center. S. B. Pandey 
is thankful to R. K. S. Yadav and J. C. Pandey for help during
observations. We are also thankful to anonymous referee for the useful comments.

\end{document}